\documentclass[12pt]{article}
\usepackage[round,comma,numbers,sort&compress]{natbib}
\usepackage{xspace}
\usepackage{times}
\usepackage{graphicx}

\newcommand{\refjnl}[1]{{\em #1\ }}
\newcommand\aj{\refjnl{Astron J.}}%
\newcommand\araa{\refjnl{Annu. Rev. Astron. Astrophys.}}%
\newcommand\apj{\refjnl{Astrophys. J.}}%
\newcommand\apjs{\refjnl{Astrophys. J. Suppl. Ser.}}%
\newcommand\aap{\refjnl{Astron. Astrophys.}}%
\newcommand\mnras{\refjnl{Mon. Not. R. Astron. Soc.}}%
\newcommand\etal{{\em et al.\xspace}}

\newenvironment{sciabstract}{%
\begin{quote} \bf}
{\end{quote}}

\newcounter{lastnote}
\newenvironment{scilastnote}{%
\setcounter{lastnote}{\value{NAT@ctr}}%
\addtocounter{lastnote}{+1}%
\begin{list}%
{\arabic{lastnote}.}
{\setlength{\leftmargin}{.22in}}
{\setlength{\labelsep}{.5em}}}
{\end{list}}

\title{Detection of a Noble Gas Molecular Ion, $^{36}$ArH$^+$, in the Crab 
Nebula}

\author{
M. J. Barlow$^{1\dag}$,
B. M. Swinyard$^{1,2}$,
P. J. Owen$^1$,
J. Cernicharo$^3$,\\
H. L. Gomez$^4$,
R. J. Ivison$^5$,
O. Krause$^6$,
T. L. Lim$^2$,\\
M. Matsuura$^1$,
S. Miller$^1$,
G. Olofsson$^7$,
E. T. Polehampton$^{2,8}$\\
\\
\normalsize{$^1$Dept. of Physics \& Astronomy, University College
London, Gower Street, London WC1E 6BT, UK}\\
\normalsize{$^2$Space Science \& Technology Department, Rutherford 
Appleton Laboratory, Didcot OX11 0QX, UK}\\
\normalsize{$^3$Laboratory of Molecular Astrophysics, Dept. of 
Astrophysics, CAB, INTA-CSIC,}\\
\normalsize{Ctra de Ajalvir, km 4, 28850 Torrej\'on de Ardoz, Madrid, Spain}\\
\normalsize{$^4$School of Physics \& Astronomy, Cardiff University, The 
Parade, Cardiff CF24 3AA, UK}\\
\normalsize{$^5$UK Astronomy Technology Centre, Royal Observatory 
Edinburgh,}\\
\normalsize{Blackford Hill, Edinburgh EH9 3HJ, UK}\\
\normalsize{$^6$Max-Planck-Institut f\"{u}r Astronomie, K\"{o}nigstuhl 
17, D-69117 Heidelberg, Germany}\\
\normalsize{$^7$Dept. of Astronomy, Stockholm University, AlbaNova 
University Center,}\\
\normalsize{Roslagstulsbacken 21, 10691 Stockholm, Sweden}\\
\normalsize{$^8$Institute for Space Imaging Science, Dept. of Physics 
\& Astronomy,}\\
\normalsize{University of Lethbridge, Lethbridge, Alberta T1K 3M4, Canada}\\
\\
\normalsize\normalsize{$^\dag$email:mjb@star.ucl.ac.uk; Published in 
Science, v.342, pp.1343-1345, 2013}
}

\date{}

\newcommand\arcsec{\mbox{$^{\prime\prime}$}\xspace}%

\begin{document}


\maketitle 


\baselineskip24pt


\begin{sciabstract} 

Noble gas molecules have not hitherto been detected in space. From 
spectra obtained with the {\em Herschel Space Observatory},
we report the detection of emission in the 617.5~GHz and 1234.6~GHz J =
1-0 and 2-1 rotational lines of $^{36}$ArH$^+$ at several positions in the
Crab Nebula, a supernova remnant known to contain both H$_2$ molecules and
regions of enhanced ionized argon emission. $^{36}$Ar 
is believed to have originated from explosive nucleosynthesis
in massive stars during core-collapse supernova events. Its detection in
the Crab Nebula, the product of such a supernova event, confirms this 
expectation. The likely excitation mechanism for the observed 
$^{36}$ArH$^+$ emission lines is electron collisions in partially 
ionized regions with electron densities of a few hundred per centimeter
cubed. 

\end{sciabstract}



Noble gas compounds have not yet been found in space, despite some 
examples, such as ionized hydrides, being relatively stable \cite{Wya75}. 
Astronomical searches for the near-infrared and far-infrared lines of 
HeH$^+$, whose dissociation energy is 1.8~eV \cite{Lia84} have not been
successful \cite{Moo88,Liu97}. The Crab Nebula is the product 
of the supernova of 1054 AD and is thought to have originated from the 
core-collapse explosion of a star 8-16 times as massive as the Sun 
\cite{Dav85}. We
have obtained far-infrared to submillimeter spectra of the Crab Nebula 
using the {\em Herschel Space Observatory} \cite{Pil10}. We report here
the detection of emission lines from the ionized hydride of argon,
an element predicted to form by explosive nucleosynthesis in
core-collapse supernovae \cite{Arn96}. 

The Crab Nebula was observed with the Fourier Transform Spectrometer
(FTS) of the Spectral and Photometric Imaging Receiver (SPIRE) 
\cite{Grif10,Swin10} on Operational Day 466 of the {\em Herschel}
mission, as part of the Mass-loss of Evolved StarS (MESS) 
Guaranteed Time Key Project \cite{Gro11}. 
The 19 SPIRE Long Wavelength (SLW) detectors, each with a beamwidth of 
$\sim$37\arcsec, covered the 447-989~GHz frequency range (303-671~$\mu$m), 
while 35 SPIRE Short Wavelength (SSW) detectors, each with a beamwidth 
of $\sim$18\arcsec, covered the 959-1544~GHz frequency range (194-313~$\mu$m)
(Fig.~1).
The full width half maximum spectral resolution was 1.44~GHz at all
frequencies, corresponding to a resolving power of 690 in the middle
of the frequency range. The observation consisted of 48 FTS scans, 
for a total on-source exposure time of 3197s. The data were processed using the
extended source calibration in version 11 of the Herschel Interactive
Processing Environment \cite{Ott10}. 
The J = 2-1, F = 5/2-3/2 line of OH$^+$ at 971.8038~GHz \cite{Mul05},
which falls in the SLW and SSW spectral overlap region, is present in
emission in many of the spectra (Fig.~2). This line has been observed from a 
range of astrophysical environments by {\em Herschel}, both in absorption
\cite{Neu10} and emission \cite{vWer10,Spin12}. In addition, two
unidentified emission lines were found to be present in some of the Crab
spectra, one in the SLW range at $\sim$618~GHz, and the other in the SSW
range at $\sim$1235~GHz. 

The knots and filaments of the Crab Nebula are known to exhibit expansion 
velocities ranging between 700-1800~km~s$^{-1}$ \cite{Hes08}; in different 
detectors we measured radial velocities for the OH$^+$ 971.8038~GHz line 
that ranged between -603 and +1037~km~s$^{-1}$. Several spectra showed 
multiple OH$^+$ velocity components, some blended, but in most spectra the 
OH$^+$ velocity components were unresolved, exhibiting very
different radial velocities from detector to detector, consistent with an 
origin from differing knots or filaments in the nebula, each with its own 
discrete velocity. Because OH$^+$ was the only identified species in the 
spectra initially, we used the measured radial velocities of the 
971.8038~GHz OH$^+$ line, whose centroid frequency could typically be 
measured 
to an accuracy of $\pm$(25-40)~km~s$^{-1}$, as a reference to correct to a 
`rest' frequency the observed frequencies of the 618 or 1235~GHz line 
falling in the same spectrum. There were four SLW spectra, those from 
detectors B3, C3, D3 and D4, in which the OH$^+$ line and the 618~GHz line 
were both present, with emission line surface brightnesses in excess of 
2$\times10^{-10}$~W~m$^{-2}$~sr$^{-1}$. These four spectra yielded a mean 
`rest' frequency for the 618~GHz line of 617.554 $\pm$ 0.209~GHz. The 
1235~GHz line was detected in five SSW spectra (A2, B1, B2, B3 and D4) but 
only in the B1 (Fig.~2) and B3 spectra were both it and the 971.8038~GHz 
OH$^+$ line present with a single unresolved component. The J = 1-1, F = 
1/2-1/2 line of OH$^+$ at 1032.998~GHz was also detected in emission in 
the SSW B1 spectrum, enabling a third estimate of the rest frequency of 
the 1235~GHz line. The mean frequency derived from these three 
estimates was 1234.786 $\pm$ 0.643~GHz.

The ratio of the derived rest frequencies of 1234.786 $\pm$ 0.643~GHz and 
617.554 $\pm$ 0.209~GHz is 1.9995$\pm$0.0012, which suggests that the 
lines correspond to the 2-1 and 1-0 rotational transitions of a simple 
diatomic molecule (we can rule out their being 4-3 and 2-1 transitions, 
with a frequency ratio of 4:2, because of the lack of a corresponding 3-2 
transition at $\sim$ 926~GHz). A search using the {\em Cologne Database 
for Molecular Spectroscopy} \cite{Mul05} and the {\sc madex} code 
\cite{Cer12} found the only candidate to be $^{36}$ArH$^+$, whose 1-0 and 
2-1 rotational transitions lie at 617.52523 $\pm$ 0.00015~GHz and 
1234.60275 $\pm$ 0.00030~GHz, respectively, agreeing with the derived 
frequencies for the Crab Nebula lines within the uncertainties. The 1-0 
and 2-1 rotational transitions of $^{40}$ArH$^+$ are at 615.85813 $\pm$ 
0.00005~GHz and 1231.27100 $\pm$ 0.00009~GHz, while the corresponding 
transitions of $^{38}$ArH$^+$ are at 616.64871 $\pm$ 0.00004~GHz and 
1232.85100 $\pm$ 0.00004~GHz, ruling out these two isotopic variants as 
identifications. Argon is the third most abundant species in the Earth's 
atmosphere, 0.93\% by number, with $^{40}$Ar/$^{38}$Ar/$^{36}$Ar isotopic 
ratios of to 1584/1.00/5.30 \cite{Lee06}. However, $^{40}$Ar in the 
earth's atmosphere is a product of the decay, mainly in rocks, of 
$^{40}$K, whose half-life is 1.25$\times10^9$~yrs. For the solar wind the 
$^{40}$Ar/$^{38}$Ar/$^{36}$Ar isotopic ratios have been measured to be 
0.00/1.00/5.50 \cite{Mes07}. $^{36}$Ar is expected to be the dominant 
isotope of argon in stars, being an explosive nucleosynthesis product of 
the $\alpha$-particle capture chain that takes place in massive star 
core-collapse supernovae \cite{Arn96}. 
Regions of enhanced emission from optical forbidden lines of 
ionized argon have previously been mapped in 
the Crab Nebula by a number of authors \cite{Mac94,Sch02,Tem12}, almost 
certainly corresponding to enriched argon abundances.
Strongly enhanced infrared forbidden lines of Ar$^+$ 
and Ar$^{2+}$ have been detected from the southern filament \cite{Tem12} 
where we find 
$^{36}$ArH$^+$ line emission to be strongest (SSW detector B1; Fig.~1 
and Table~1). The enhanced lines of ionized argon and $^{36}$ArH$^+$ 
strongly indicate the presence there of pockets of $^{36}$Ar produced by 
explosive nucleosynthesis during the supernova event.

The $^{38}$Ar isotope of argon is also predicted to be synthesised in 
core-collapse supernova events \cite{Arn96}. We used Crab Nebula 
SPIRE-FTS spectra in which the $^{36}$ArH$^+$ velocity components are 
strong and narrow to put limits on the isotopic ratios of $^{40}$Ar and 
$^{38}$Ar relative to $^{36}$Ar in the emitting regions. Because the 
frequency separation between the isotopic variants of ArH$^+$ is a factor 
of two larger for the J = 2-1 lines than for the 1-0 lines, whereas the 
SPIRE-FTS frequency resolution is constant, the 2-1 lines are better 
suited for placing limits on the isotopic ratios of argon. Relative to the 
$^{36}$ArH$^+$ 2-1 line, the separation of the $^{40}$ArH$^+$ 2-1 line is
-3.33175~GHz, while that of the $^{38}$ArH$^+$ 2-1 line is -1.75175~GHz. 
To estimate upper limits to the abundances of these species, we added 
synthetic lines to the spectra having the appropriate frequency offset 
and the same line width as the $^{36}$ArH$^+$ 2-1 line, changing the 
strength of the synthetic lines until the signal-to-noise ratio 
estimated by the line-fitting routine reached 3$\sigma$.
The SSW~B1 (Fig.~2) and SSW~B3 spectra both yielded 3$\sigma$ lower limits 
to the abundance ratio of $^{36}$ArH$^+$/$^{38}$ArH$^+$ of $>2$, along
with $^{36}$ArH$^+$/$^{40}$ArH$^+$ lower limits of $>5$ and $>4$, 
respectively. 

ArH$^+$ is a stable molecular ion (dissociation energy D$_0$ = 
3.9$\pm$0.1~eV \cite{Wya75}) that has been studied extensively in the 
laboratory. The Crab 
Nebula consists predominantly of ionized gas, photoionized by synchrotron 
radiation from the pulsar wind nebula \cite{Dav85, Hes08}. It also contains many 
H$_2$-emitting neutral clumps \cite{Gra90,Loh11}. Transition zones between 
fully ionized and molecular gas will exist, where ArH$^+$ can be formed by 
the exothermic reaction Ar$^+$ + H$_2$ $\rightarrow$ ArH$^+$ + H, 
releasing 1.49~eV \cite{Roa70}. If the elemental species created by the 
supernova explosion were still largely unmixed in the remnant, then it is 
possible that ArH$^+$ molecules would be found only at interfaces between 
H-rich gas and Ar-rich gas where mixing has occurred. Four of the seven 
FTS SSW detectors in whose spectra J~=~2-1 $^{36}$ArH$^+$ emission was 
detected (SSW B1, B2, A1 and A2) are situated on a bright filament south 
of the center of the nebula (Fig.~1), as is the SLW D4 detector in which 
the strongest J~=~1-0 emission was detected. A cluster of seven 
near-infrared H$_2$-emitting knots, with a wide range of radial 
velocities, is coincident with the same bright filament \cite{Loh11}. 
Detectors SLW~C3/SSW~D4 also show $^{36}$ArH$^+$ emission and are 
coincident with an H$_2$-emitting knot. The lack of ArH$^+$ emission in 
the NW quadrant of the nebula is mirrored by a relative lack of H$_2$ 
emission knots in that region. However, there are many H$_2$ knots in the 
NE quadrant, whereas ArH$^+$ emission is only detected there in two FTS 
detectors (SLW~E1 and E2).

The reaction rate for the formation of ArH$^+$ via Ar$^+$ + H$_2$
$\rightarrow$ ArH$^+$ + H is $8.9\times10^{-10}$~cm$^3$~s$^{-1}$ at 300~K
\cite{Roa70}. Reaction rates are known for the ArH$^+$ destruction
reaction with H$_2$ (to yield Ar + H$_3^+$, with a reaction rate of
$6.3\times10^{-10}$~cm$^3$~s$^{-1}$ at 300~K \cite{Ani03}) and for its
dissociative recombination with electrons (ArH$^+$ + e$^-$ $\rightarrow$
Ar + H$^*$, with a rate $\leq 5\times10^{-10}$~cm$^3$~s$^{-1}$ at low
electron energies \cite{Mit05}). However, because the Crab nebula is
photoionized by its pulsar wind nebula \cite{Dav85, Hes08},
photodissociation could be the main ArH$^+$ destruction mechanism.

The main excitation mechanism for the observed ArH$^+$ emission lines is 
likely to be collisions with either electrons or H$_2$ molecules but rate 
calculations or measurements do not exist as yet. The J~=~1 and 2 levels 
of $^{36}$ArH$^+$ are situated 29.6~K and 88.9~K above the ground state, 
respectively, negligible compared to the electron temperatures of 
7500-15000~K measured for the ionized gas in the Crab Nebula \cite{Dav85}, 
or even compared to the H$_2$ excitation temperatures of 2000-3000~K that 
have been measured \cite{Loh12}. If the electron or H$_2$ densities in the 
transition zones where ArH$^+$ is hypothesised to be located should exceed 
the `critical densities' of the emitting levels (where the sum of the 
collisional excitation and de-excitation rates from a level exceed the 
radiative decay rate from the level), then the level populations will be 
in Boltzmann equilibrium. Using the known molecular parameters of ArH$^+$ 
\citep{Mul05, Cer12} the 2-1/1-0 line emission ratios should then be of 
the order 30, for excitation temperatures appreciably exceeding 100~K. 
The SSW~D4 and SLW~C3 detectors are centred on the same bright knot 
(Fig.~1) and yield a 2-1/1-0 line surface brightness ratio of 2.5, while 
the spectra from the approximately co-located SSW~B3 and SLW~C4 detectors 
yield a line surface brightness ratio of 2.0 (Table~1), well below the 
ratio for Boltzmann equilibrium. The densities of the collision partners 
in the emitting regions must therefore be well below the 
corresponding critical densities of the ArH$^+$ rotational levels. We used 
the {\sc madex} code \cite{Cer12} with the molecular parameters of 
ArH$^+$, together with SiH$^+$ + He collisional de-excitation rates 
\cite{Nkem09} in place of those of ArH$^+$ + H$_2$, and CH$^+$ + e$^-$ 
collisional de-excitation rates \cite{Lim99} in place of those for ArH$^+$ 
+ e$^-$ (with upward rates calculated using detailed balance and the 
correct values of the energies for the levels of ArH$^+$) in order to 
estimate corresponding H$_2$ and electron critical densities of 
$\sim10^8$~cm$^{-3}$ and $\sim10^4$~cm$^{-3}$, respectively. The 
observed 2-1/1-0 line ratios of 2.5 and 2.0 indicate H$_2$ densities of 
a few$\times10^6$~cm$^{-3}$, or electron densities of a 
few$\times10^2$~cm$^{-3}$. The calculations take into account opacity 
effects although the line centre optical depths are estimated to be 
significantly less than unity for line widths larger than 1~km~s$^{-1}$.
For H$_2$ collisions at temperatures between 100~K and 3000~K and 
densities of $\sim10^6$~cm$^{-3}$, or electron collisions at temperatures
of $\sim$3000~K and densities of a few$\times10^2$~cm$^{-3}$,
we estimate $^{36}$ArH$^+$ column densities of $10^{12}-10^{13}$~cm$^{-2}$.

Given that likely electron collisional excitation rates are $\sim10^4$ 
larger than those of H$_2$ and that the parent Ar$^+$ ion must exist in a 
region that is at least partially ionized, electron collisions are 
expected to dominate the excitation of ArH$^+$. A density of 
$\sim10^4$~cm$^{-3}$ has been estimated for the Crab H$_2$ knots 
\cite{Loh12}, a factor of 100 below that required for H$_2$ collisions to 
be the excitation mechanism for the ArH$^+$ lines. This lends further 
support to electron collisions in partially ionized transition regions 
being the main ArH$^+$ excitation mechanism. 

Our detection of $^{36}$ArH$^+$ in the Crab Nebula suggests that an unidentified
multi-component broad absorption feature seen between 617 and 618~GHz in a
Herschel HIFI spectrum of Sgr B2(M) (towards the center of our galaxy)
\cite{Sch10} can potentially be identified with ground-state absorption 
in the J = 0-1 617.525 GHz line of $^{36}$ArH$^+$ along the interstellar 
sightline.



\begin{thebibliography}{}

\bibitem[Wyatt \etal (1975)]{Wya75} J.~R. Wyatt \etal, 
    {\em J. Chem. Phys.} {\bf 62}, 2555 (1975).

\bibitem[Lias \etal (1984)]{Lia84} S.~G. Lias, J.~F. Liebman, R.~D. Levin, 
  {\em J. Phys. Chem. Ref. Data} {\bf 13}, 695 (1984).

\bibitem[Moorhead \etal (1998)]{Moo88} J.~M. Moorhead,
 R. P. Lowe, W. H. Wehlau, J.-P. Maillard, P. F. Bernath,
  \apj {\bf 326}, 899 (1988). 

\bibitem[Liu \etal (1997)]{Liu97} X.-W. Liu \etal,
  \mnras {\bf 290}, L71 (1997).

\bibitem[Davidson \& Fesen (1985)]{Dav85} K. Davidson, R. A. Fesen,
    \araa {\bf 23}, 119 (1985).

\bibitem[Pilbratt \etal (2010)]{Pil10} G.~L. Pilbratt \etal,
  \aap {\bf 518}, L1 (2010).

\bibitem[Arnett (1996)]{Arn96} D. Arnett, 
   {\em Supernovae and Nucleosynthesis}, (Princeton Univ. Press, 1996).
  
\bibitem[Griffin \etal (2010)]{Grif10} M.~J. Griffin \etal,
  \aap {\bf 518}, L3 (2010).

\bibitem[Swinyard \etal (2010)]{Swin10} B.~M. Swinyard \etal,
  \aap {\bf 518}, L4 (2010).

\bibitem[Groenewegen \etal (2011)]{Gro11} M. Groenewegen \etal,
  \aap {\bf 526}, A162 (2011).

\bibitem[Ott \etal (2010)]{Ott10} S. Ott,
  {\em Astr. Soc. Pacific Conf. Ser.} {\bf 434}, 139 (2010).

\bibitem[Muller \etal (2005)]{Mul05} H.~S.~P. M\"{u}ller,
  F. Schl\"{o}der, J. Stutzki, G. Winnewisser,
  {\em J. Mol. Struct.} {\bf 742}, 215 (2005).

\bibitem[Neufeld \etal (2010)]{Neu10} D.~A. Neufeld \etal,
   \aap {\bf 521}, L10 (2010).

\bibitem[van der Werf \etal (2010)]{vWer10} P.~P. van der Werf \etal,
  \aap {\bf 518}, L42 (2010).

\bibitem[Spinoglio \etal (2012)]{Spin12} L. Spinoglio \etal,
   \apj {\bf 758}, 108 (2012).

\bibitem[Hester \etal (2012)]{Hes08} J.~J. Hester,
  \araa {\bf 46}, 127 (2008).

\bibitem[Cernicharo (2012)]{Cer12} J. Cernicharo,
   {\em EAS Publ. Ser.}  {\bf 58} 251 (2012).

\bibitem[Lee \etal (2012)]{Lee06} J.-Y. Lee \etal,
   {\em Geochim. et Cosmochim. Acta} {\bf 70}, 4507 (2006).

\bibitem[Meshik \etal (2007)]{Mes07} A. Meshik \etal,
  {\em Science} {\bf 318}, 433 (2007).

\bibitem[MacAlpine \etal (1994)]{Mac94} G.~M. MacAlpine \etal,
     \apj {\bf 432}, L131 (1994).

\bibitem[Schaller \& Fesen (2002)]{Sch02} E.~L. Schaller \& R.~A. Fesen, 
     \aj {\bf 123}, 941 (2002).

\bibitem[Temim \etal (2012)]{Tem12} T. Temim \etal,
    \apj {\bf 753}, 72 (2012).



\bibitem[Graham \etal (1990)]{Gra90} J.~R. Graham, G.~S. Wright, A.~J. Longmore,
    \apj {\bf 352}, 172 (1990).

\bibitem[Loh \etal (2011)]{Loh11} E.~F.~Loh \etal,
     \apjs {\bf 194}, 30 (2011).

\bibitem[Roach \etal (1970)]{Roa70} A.~C. Roach, P.~J. Kuntz,
      {\em Chem. Comms.} 1336 (1970).

\bibitem[Anicich (2003)]{Ani03} V.~G. Anicich,
  {\em NASA JPL Publ.} 03-19 (2003).


\bibitem[Mitchell \etal (2005)]{Mit05} J.~B.~A. Mitchell \etal,
   {\em J. Phys. B} {\bf 38}, L175 (2005).

\bibitem[Loh \etal (2012)]{Loh12} E.~D. Loh \etal,
    \mnras {\bf 421}, 789 (2012).

\bibitem[Nkem \etal (2009)]{Nkem09} C. Nkem \etal,
   {\em J. Mol. Struct. \sc theochem} {\bf 901}, 220 (2009).

\bibitem[Lim \etal (1999)]{Lim99} A.~J. Lim, I. Rabad\'{a}n, J. Tennyson,
   \mnras {\bf 306}, 473 (1999). 

\vspace{0.4cm}

\bibitem[Schilke \& HEXOS Team (2010)]{Sch10} P. Schilke, HEXOS Team,
   in {\em Proceedings of the Herschel First Results Symposium}, (2010).
{\scriptsize
\begin{verbatim}
http://herschel.esac.esa.int/FirstResultsSymposium/presentations/A34_SchilkeP_SgrB2.pdf   
\end{verbatim}} 

\bibitem[Gomez \etal (2012)]{Gom12} H.~L. Gomez \etal,
   \apj {\bf 760}, 96 (2012).


\end{thebibliography}


\begin{scilastnote}
\item 
We thank the anonymous referees for their constructive reports.
We thank Dr Edwin Bergin for drawing our attention to the 2010 HEXOS 
Herschel HIFI spectrum of SgR B2(M) that shows
an unidentified broad absorption feature below 618~GHz.
{\it Herschel} is an ESA space observatory with science 
instruments provided by European-led Principal Investigator consortia and 
with important participation from NASA.
SPIRE has been developed by a consortium of institutes led by Cardiff
University (UK) and including Univ. Lethbridge (Canada); NAOC (China);
CEA, LAM (France); IFSI, Univ. Padua (Italy); IAC (Spain); Stockholm
Observatory (Sweden); Imperial College London, RAL, UCL-MSSL, UKATC, Univ.
Sussex (UK); and Caltech, JPL, NHSC, Univ. Colorado (USA). This
development has been supported by national funding agencies: CSA (Canada);
NAOC (China); CEA, CNES, CNRS (France); ASI (Italy); MINECO (Spain); SNSB
(Sweden); STFC and UKSA (UK); and NASA (USA). JC thanks MINECO for 
financial support under projects AYA2009-07304, AYA2012-32032 and 
Consolider program ASTROMOL CSD2009-00038.

\end{scilastnote}


\clearpage

\begin{table}\centering
\caption{SPIRE-FTS radial velocity and line surface brightness 
measurements for the J = 1-0 and 2-1 rotational lines of $^{36}$ArH$^+$ 
from the Crab Nebula}
\begin{tabular}{c c c c c c} \\ \hline
\multicolumn{3}{c}{J = 1-0~~~ 617.525~GHz} & \multicolumn{3}{c}{J = 2-1~~~ 1234.603~GHz}  \\
 SLW       & Radial Velocity  &  Surface Brightness      & SSW   & Radial Velocity &  Surface Brightness \\
 Detector  &  km~s$^{-1}$  & $10^{-10}$~W~m$^{-2}$~sr$^{-1}$  & Detector  &  km~s$^{-1}$ & $10^{-10}$~ W~m$^{-2}$~sr$^{-1}$ \\
\hline
  B3       &  +317 $\pm$ 67 &  2.23 $\pm$ 0.41  &  C5   & -1354 $\pm$ 26 &  8.2  $\pm$ 1.2  \\
  C3       &  +933 $\pm$ 33 &  4.63 $\pm$ 0.40  &  D4   &  +743 $\pm$ 26 &  11.7 $\pm$ 1.6  \\
  C4       &  -58  $\pm$ 50 &  8.65 $\pm$ 0.55  &  B3   &  -101 $\pm$ 20 &  17.5 $\pm$ 1.4  \\ 
  D3       &  +826 $\pm$ 32 &  3.13 $\pm$ 0.34  &       &                &                  \\
  D3       &  -709 $\pm$ 42 &  2.30 $\pm$ 0.34  &       &                &                  \\
  D4       &  +101 $\pm$ 27 &  9.89 $\pm$ 0.52  &  A1   &   -51 $\pm$ 52 &  13.9 $\pm$ 2.0  \\
           &                &                   &  B2   &  -572 $\pm$ 25 &  10.8 $\pm$ 1.7  \\
           &                &                   &  B1   &  +140 $\pm$ 34 &  38.4 $\pm$ 1.6  \\
           &                &                   &  A2   &   +61 $\pm$ 28 &  10.1 $\pm$ 1.4  \\
  E1       &  +278 $\pm$ 46 &  5.69 $\pm$ 0.62  &       &                &                  \\
  E2       &  -594 $\pm$ 37 &  4.25 $\pm$ 0.46  &       &                &                  \\
\\ \hline
\end{tabular}
\end{table}

\clearpage

{\bf Fig. 1.}
A broad-band Herschel-PACS image mapping the 70-$\mu$m 
dust emission from the Crab Nebula. 
North is up and East is to the left.
The positions on the nebula of the 19 SLW and 35 SSW detectors
of the SPIRE FTS are marked with circles whose angular 
diameters of 37\arcsec and 18\arcsec correspond to the 
SLW and SSW beam sizes. The positions of detectors in whose spectra the J 
= 1-0 or 2-1 rotational lines of $^{36}$ArH$^+$ were detected are marked 
with crosses. \\

{\bf Fig. 2.} 
SPIRE FTS spectra of the Crab Nebula, plotting surface brightness
in W~m$^{-2}$~Hz$^{-1}$~sr$^{-1}$ against frequency in GHz. 
Several emission lines are superposed on a continuum attributed to
thermal dust emission \cite{Gom12}. Upper plot: the spectrum from
the SLW D4 detector, whose position on the nebula is marked in Fig.~1.  
Emission line velocity components attributed to the J = 2-1, F = 5/2-3/2 
971.8038~GHz rotational line of OH$^+$ and the J = 1-0  617.525~GHz
rotational line of $^{36}$ArH$^+$ are visible.
Lower plot: the SSW B1 spectrum. In addition to OH$^+$ 971.8038~GHz 
velocity components, emission in the J = 2-1 1234.603~GHz 
line of $^{36}$ArH$^+$ is visible. The radial velocities 
and surface brightnesses of the $^{36}$ArH$^+$ emission
lines that are present in the spectra obtained from these and other FTS
detectors are listed in Table~1.

\begin{figure}
\begin{center}
\includegraphics[width=6.5in]{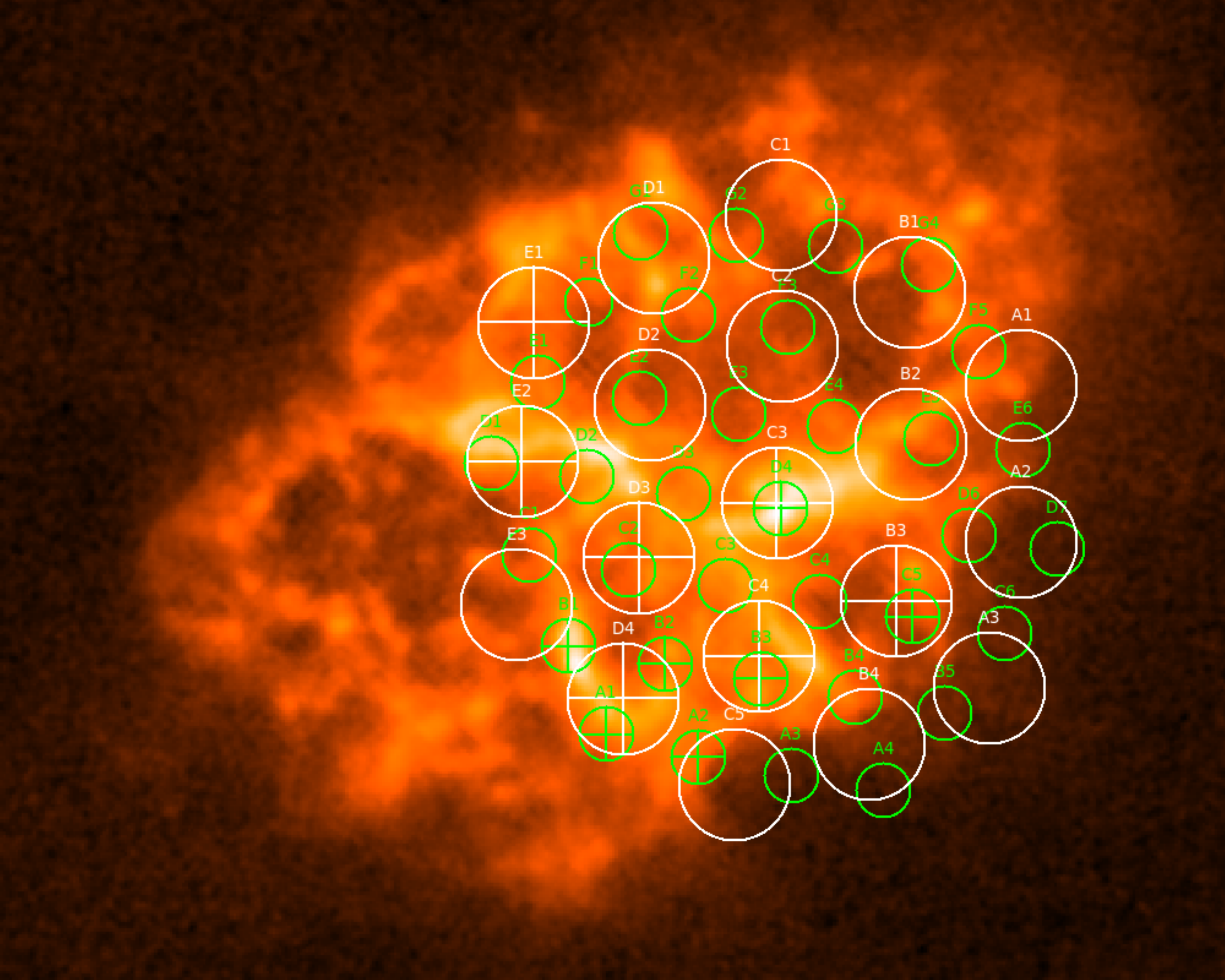}
\end{center}
\end{figure}

\begin{figure}
\begin{center}
\includegraphics[width=6.5in]{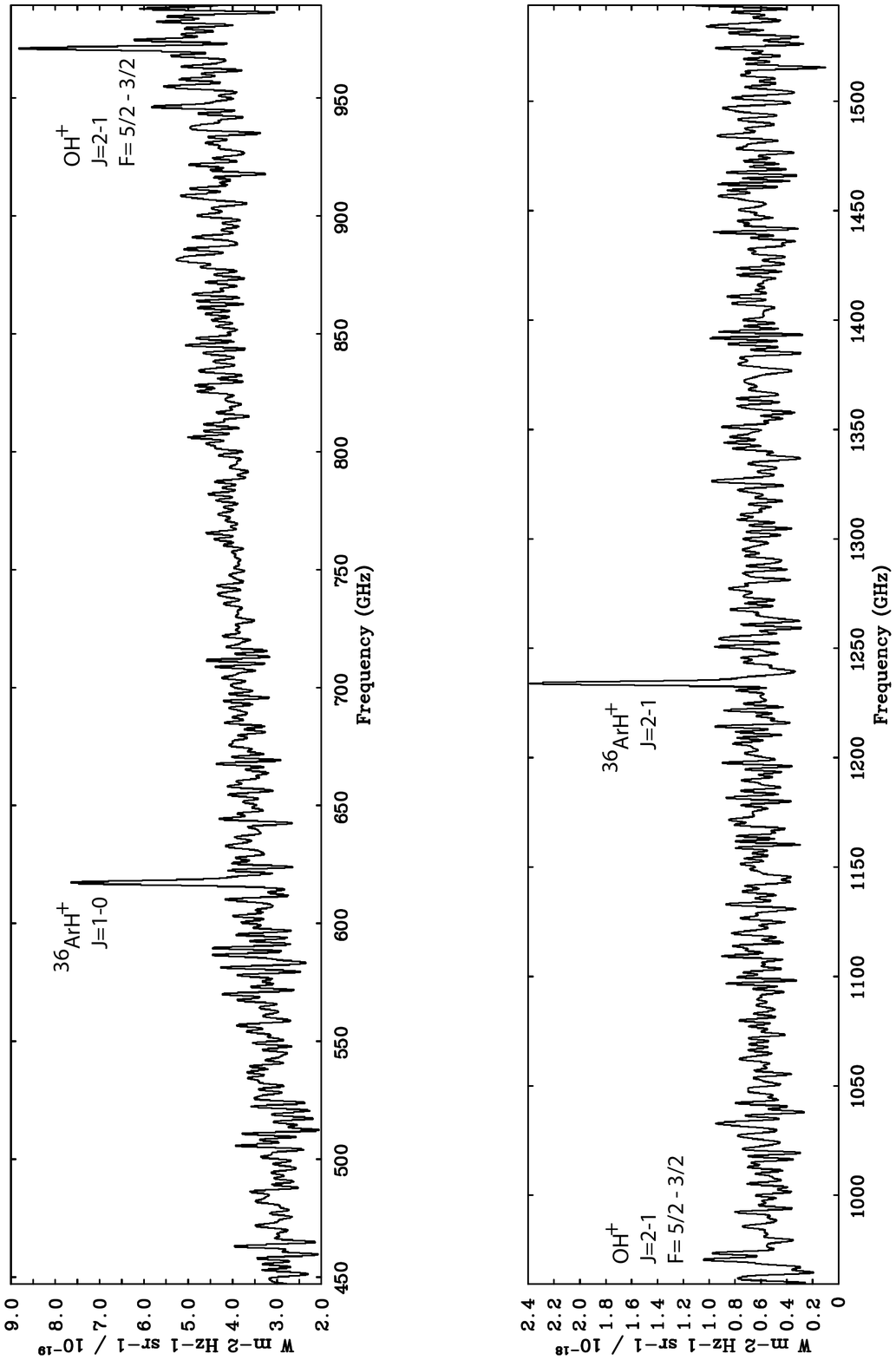}
\end{center}
\end{figure}

\end{document}